\documentclass[12pt]{article}
\usepackage{amsmath,epsfig,amssymb,epstopdf}
\newcommand{\be}{\begin{equation}}
\newcommand{\ee}{\end{equation}}
\newcommand{\bear}{\begin{eqnarray}}
\newcommand{\eear}{\end{eqnarray}}
\newcommand{\ba}{\begin{array}}
\newcommand{\ea}{\end{array}}

\begin{document}

\begin{flushright}
{\tt hep-th/yymmnnn}
\end{flushright}

\vspace{5mm}

\begin{center}
{{\Large \bf The holographic superconductors \\in higher-dimensional AdS soliton}\\[14mm]
{Chong Oh Lee}\\[2.5mm]
{\it Department of Physics, Kunsan National University,\\
Kunsan 573-701, Korea}\\
{\tt cohlee@kunsan.ac.kr}}
\end{center}

\vspace{10mm}

\begin{abstract}
We explore the behavior of the holographic superconductors at zero temperature for a charged scalar
field coupled to a Maxwell field in higher-dimensional AdS soliton spacetime via
analytical way. In the probe limit, we obtain the critical chemical potentials increase linearly
as a total dimension $d$ grows up.
We find that the critical exponent for condensation operator is obtained as 1/2 independently of $d$,
and the charge density is linearly related to the chemical potential near the critical point.
Furthermore, we consider a slightly generalized setup the Einstein-Power-Maxwell field theory, and find
that the critical exponent for condensation operator is given as $1/(4-2n)$ in terms of a power parameter
$n$ of the Power-Maxwell field, and the charge density is proportional to the chemical potential
to the power of $1/(2-n)$.
\end{abstract}

\newpage
\setcounter{equation}{0}
\section{Introduction}
Nonlinear theory of electrodynamics has been suggested in Ref. \cite{Born:1934aa} in search for
an improvement over Maxwell theory with a infinite electrostatic self-energy of a point,
and its extended form has been obtained in Ref. \cite{Hoffmann:1935ty}. It has been found
in Ref. \cite{Heisenberg:1935qt} through investigation of transition to state of virtual charged
particle in quantum electrodynamics. It has been also studied in gravity theory. For example,
black hole solutions are obtained from nonlinear
electrodynamics minimally coupled to gravity for a static and spherical symmetric spacetime
\cite{deOliveira:1994in}, and by nonlinear electrodynamics with power-law function
\cite{Hassaine:2007py}.

On the other hand, for asymptotically AdS spacetime, it is of interest to attempt to study
the phase transition in the model for holographic superconductors \cite{Gubser:2008px, Hartnoll:2008vx}
since it allows new predictions through exploring the proposed AdS/CFT
correspondence \cite{Maldacena:1997re, Gubser:1998bc, Witten:1998qj},
which relates a gravitational theory on asymptotically
in the bulk to a conformal field theory in the boundary. Their behavior
has been explored by a gravitational theory of a charged scalar field coupled to a Maxwell field
\cite{Hartnoll:2009sz, Herzog:2009xv, Horowitz:2010gk}.
The gravity model of the holographic superconductor has revived many investigations for their potential
applications along these directions \cite{Nakano:2008xc}-\cite{Brihaye:2010mr}. A few phase
transition studies in a Stueckelberg form have been carried out \cite{Franco:2009yz}-\cite{Gangopadhyay:2012am}.
Furthermore a superconducting phase dual to the AdS soliton configuration is interesting case
\cite{Nishioka:2009zj}-\cite{Cai:2011tm}
since the AdS black hole in the Poincar$\rm{\acute{e}}$ coordinate can exhibit a phase transition
to the AdS soliton even if the AdS black hole and the AdS soliton have the same boundary topology
in asymptotically AdS spacetimes \cite{Surya:2001vj}.

Even if the model for holographic superconductors is well established in four- and five-dimensional spacetime
it is less explored in higher-dimensional spacetime.
Thus, one intriguing question is that of the higher-dimensional behavior for holographic superconductors.
Another is how they are affected from the Power-Maxwell field
since they are governed by the gravity theory with electric field coupled to the charged scalar field.

In this paper we consider the Einstein-Maxwell field theory in higher-dimensional AdS soliton
and find the critical exponent for condensation operator is 1/2 independently of $d$ in the limit of probe
at zero temperature, and the charge density is directly proportional to the chemical potential.

The paper is organized as follows: In the next section we investigate
the model for holographic superconductors.
We obtain the critical chemical potentials for various dimensions
of operators in $d$-dimensional spacetime, and the relations
between the charge density and the chemical potential near the critical point.
In the last section we give our conclusion.

\setcounter{equation}{0}
\section{Holographic Duality in the AdS soliton background}
In this section, we will construct the phase transition model for the Einstein-Power-Maxwell
field theory in the AdS soliton background.

Considering a superconductor dual to a AdS soliton configuration in the probe limit,
the line element of $d$-dimensional AdS soliton is given by \cite{Nishioka:2009zj,Witten:1998zw, Horowitz:1998ha}
\bear\label{metric1}
ds^2=\frac{dr^2}{f(r)}+\frac{r^2}{L^2}(-dt^2+h_{ij}dx^i dx^j)+f(r)d\eta^2,
\eear
with
\bear
f(r)=\frac{r^2}{L^2}\left(1-\frac{L^{d-1}r_0^{d-1}}{r^{d-1}}\right),
\eear
where $L$ is AdS radius and $r_0$ is the tip of soliton.
One must impose the periodicity $\eta\sim \eta+\frac{\pi}{r_0}$ to avoid a conical
singularity~\cite{Clarkson:2005qx}.
The $d$-dimensional Power-Maxwell-scalar action with negative cosmological constant is
\bear
S=\int d^dx\sqrt{-g}&&\bigg\{R-2\Lambda-\alpha(F_{\mu\nu}F^{\mu\nu})^n
-\partial_{\mu}\Psi \partial^{\mu}\Psi-m^2\Psi^2\\\nonumber
&&-\Psi^2(\partial_{\mu}\Phi-qA_{\mu})(\partial^{\mu}\Phi-qA^{\mu})
\bigg\},
\eear
where $g$ denotes the determinant of the metric, $R$ the Ricci scalar, and $\Lambda=(d-1)(d-2)/L^2$
the cosmological constant. $F^{\mu\nu}$ is the strength of the Power-Maxwell (PM) field
$F=dA$, the complex scalar field $\Psi$, the coupling constant $\alpha$, and the power of PM field $n$.
We may take the solutions of $r$ only,
\bear
A=\phi(r)dt,~~~~~~\Psi=|\Psi|=\psi(r),
\eear
and impose the gauge choice $\Phi=0$, and set $L=1$ and $q=1$
through appropriately scaling symmetries in as \cite{Gregory:2009fj}.
Then the equations of motion are given by
\bear
\ddot{\psi}+\left(\frac{\dot{f}}{f}+\frac{d-2}{r}\right)\dot{\psi}
+\left(\frac{r^2\phi^2}{f}-\frac{m^2}{f}\right)\psi=0,\\
\ddot{\phi}+\left\{\frac{\dot{f}}{f}+
\left(\frac{d-4}{2n-1}\right)\frac{1}{r}\right\}\dot{\phi}+\frac{1}{\alpha n(2n-1)(-2)^{n}}
\frac{\psi^2\phi}{\dot{\phi}^{2(n-1)}f}=0,
\eear
which leads to
\bear\label{fieldeq3}
\psi^{''}+\left(\frac{f^{'}}{f}-\frac{d-4}{z}\right)\psi^{'}
+\frac{r_0^2}{z^4}\left(\frac{z^2\phi^2}{f}-\frac{m^2}{f}\right)\psi=0,\\
\label{fieldeq4}
\phi^{''}+
\left\{\frac{f^{'}}{f}-\left(\frac{d-4}{2n-1}-2\right)\frac{1}{z}\right\}\phi^{'}
-\frac{r_0^{2n}}{\alpha n(2n-1)(-1)^{3n+1}2^nz^{4n}}\frac{\psi^2\phi}{(\phi^{'})^{2(n-1)}f}=0,
\eear
by introducing a new coordinate $z=r_0/r$.
Here a dot denotes the derivative with respect to $r$ and
a prime is the derivative with respect to $z$.

In order to solve the above equations,
one needs to impose boundary condition at the tip $z=1$ ($r=r_0$) and one at the origin $z=0$ ($r=\infty$).
Thus, at the tip one can do the expansion
\bear\label{bc1}
\psi(z)&=&a_1+a_2(z-1)+a_3(z-1)^2+\cdots,\\
\label{bc2}
\phi(z)&=&b_1+b_2(z-1)+b_3(z-1)^2+\cdots,\\
\label{bc3}
f(z)&=&c_2(z-1)+\cdots,
\eear
whose solutions behave as
\bear\label{bc11}
\psi(z=1)&=&a_1,\\
\label{bc22}
\phi(z=1)&=&b_1,
\eear
where $a_1$ and $b_1$ are constants.
Since one can set $r_0=1$
through appropriately scaling symmetries in as \cite{Gregory:2009fj}, at the origin, the solutions behave as
\bear
\label{bc4}
\psi&=&z^{\lambda_{-}}\,\psi_{-}+z^{\lambda_{+}}\,\psi_{+},\\
\label{bc5}
\phi&=&\mu-\rho z^{(d-2)/(2n-1)-1},
\eear
with
\bear\label{lam1}
\lambda_{\pm}=\frac{1}{2}\left\{(d-1)\pm\sqrt{(d-1)^2+4m^2}\right\},
\eear
and hereafter $r_0=1$.
In light of AdS/CFT correspondence,
$\psi_\pm$ can be interpreted as the expectation value of the operator $\cal{O}_{\pm}$
dual to the charged scalar field $\psi$
\bear
\label{bc44}
\psi&=&z^{\lambda_{-}}\,<{\cal O}_{-}>
+z^{\lambda_{+}}\,<{\cal O}_{+}>,
\eear
and the constants $\mu$ and $\rho$ are able to be considered
as the chemical potential and charge density in the dual field theory.
Since the condensation goes to zero ($\psi\rightarrow 0$) near the critical temperature,
the Eq. (\ref{fieldeq4}) reduces to
\bear\label{gosol1}
\phi^{''}+\left\{\frac{f^{'}}{f}-\left(\frac{d-4}{2n-1}-2\right)\frac{1}{z}\right\}\phi^{'}=0,
\eear
which yields the general solution
\bear
\phi=\beta+\gamma g(z),
\eear
whose integration constants $\beta$ and $\gamma$
are determined by the boundary conditions (\ref{bc11}), (\ref{bc22}), (\ref{bc4}), and (\ref{bc5})
\bear\label{gensolga1}
\phi=\mu,
\eear
i.e. in order to render the gauge field finite near the tip,
the Neumann boundary condition near $z=1$ imposes $\gamma=0$
so that $\beta$ is obtained as $\mu$.
This means $\phi$ has only constant solution independent of the power of the Power-Maxwell field $n$
for any dimension $d$
in as the Einstein-Maxwell-scalar theory \cite{Cai:2011ky}.
Near the origin $z=0$, one can introduce a trial function $F(z)$ for $\psi(z)$ as in \cite{Siopsis:2010uq}
\bear\label{refield1}
\psi(z)|_{z\rightarrow 0}\sim<{\cal O}_{\pm}>\,z^{\lambda_{\pm}}\,F(z),
\eear
which satisfies $F(0)=1$ and $F^{'}(0)=0$. Substituting Eqs. (\ref{gensolga1})
and (\ref{refield1}) into Eq. (\ref{fieldeq3}) we get
\bear
F^{''}(z)&+&\left\{-\frac{(d-2)z^{d-1}+2}{z(1-z^{d-1})}
+\frac{2\lambda_{\pm}}{z}-\frac{d-4}{z}\right\}F^{'}(z)\\\nonumber
&+&\left[\frac{\lambda_{\pm}(\lambda_{\pm}-1)}{z^2}-\frac{\lambda_{\pm}}{z}
\left\{\frac{(d-3)z^{d-1}+2}{z(1-z^{d-1})}+\frac{d-4}{z}\right\}\right.\\\nonumber
&+&\left.\frac{\mu^2}{1-z^{d-1}}-\frac{m^2}{z^2(1-z^{d-1})}\right]F(z)=0,
\eear
which leads to
\bear
\left\{T(z)F^{'}(z)\right\}^{'}-P(z)F(z)+\mu^2Q(z)F(z)=0,
\eear
via the following functions:
\bear
T(z) &=& z^{2\lambda_{\pm}-3}(z^{d-1}-1),\\\nonumber
P(z) &=& -T(z)\left[\frac{\lambda_{\pm}(\lambda_{\pm}-1)}{z^2}-\frac{\lambda_{\pm}}{z}
\left\{\frac{(d-3)z^{d-1}+2}{z(1-z^{d-1})}+\frac{d-4}{z}\right\}
-\frac{m^2}{z^2(1-z^{d-1})}\right],\\\nonumber
Q(z) &=& \frac{T(z)}{1-z^{d-1}}
\eear
After setting the trial function $F(z)=1-az^2$, the minimum eigenvalues of $\mu^2$ is calculated
from the variation of the following functional \cite{Siopsis:2010uq}
\bear\label{camipot}
\mu^2=\frac{\int_{0}^{1}dz\bigg\{T(z)F^{'2}(z)+P(z)F^2(z)\bigg\}}{\int_{0}^{1}dz Q(z) F^2(z)}.
\eear
After taking $m^2=d(d-2)/4$, from Eq.(\ref{lam1})
we get the operator ${\cal O}_{-}$ of conformal dimension
\bear
\lambda_{-}=\frac{d-2}{2},
\eear
Then, $\mu_{-}^2$ is explicitly given by
\bear
\mu_{-}^2=\frac{s_{\mu_{-}}(a,d)}{t_{\mu_{-}}(a,d)},
\eear
where
\bear
s_{\mu_{-}}(a,d) &=& d(d-4) \bigg\{(2d-5)(2d-7)(d^3-6d^2+28d-24)a^2\\\nonumber
&&-2(d-2)^3(2d-7)(2d-3)a+(d-2)^3(2d-5)(2d-3)\bigg\},\\\nonumber
t_{\mu_{-}}(a,d) &=& 4 (2d-3) (2d-5) (2d-7) \bigg\{(d-2)(d-4)a^2-2d(d-4)a+d(d-2)\bigg\}.
\eear
When the constant $a_{-}$ is
\bear
a_{-}=\frac{s_{a_{-}}(d)}{t_{a_{-}}(d)},
\eear
\bear
s_{a_{-}}(d) &=& 2d^6-11d^5+7d^4+12d^3+132d^2-376d+240-2\bigg(53d^{10}-882d^9\\\nonumber
&&+6094d^8-22310d^7+44985d^6-43972d^5+5624d^4+16608d^3+12448d^2\\\nonumber
&&-33024d+14400\bigg)^{1/2},\\\nonumber
t_{a_{-}}(d) &=& 2d^6-d^5-129d^4+578d^3-620d^2-472d+672,
\eear
the minimum eigenvalue $\mu_{\rm min(-)}$ yields
\bear\label{cofmin1}
\mu_{\rm min(-)} = \frac{s_{\mu_{\rm min(-)}}(d)}{t_{\mu_{\rm min(-)}}(d)},
\eear
with
\bear
s_{\mu_{\rm min(-)}}(d) &=& \Bigg\{11d^5-105d^4+371d^3-600d^2+440d-120
-(d-2)\bigg (53d^8\\\nonumber
&&-670d^7+3202d^6-6822d^5+4889d^4+2872d^3-2444d^2-4656d+3600\bigg)^{1/2}\Bigg\}^{1/2},\\\nonumber
t_{\mu_{\rm min(-)}}(d) &=& 2\bigg( (2d-3) (2d-5) (2d-7)\bigg)^{1/2}.
\eear
For example, the minimum eigenvalue $\mu_{\rm min}$ (\ref{cofmin1}) for $d=5$ is given by
$\mu_{c}=\mu_{\rm min(-)}\thickapprox 0.837$,
which is exactly matched with that in \cite{Cai:2011ky},
and $\mu_{\rm min(-)}\thickapprox 1.22$ for $d=6$, and $\mu_{\rm min(-)}\thickapprox 1.58$ for $d=7$.

When the scalar field squared mass $m^2$ is bigger than the Breitenlohner-Freedman bound
squared mass $m_{\rm BF}^2=-(d-1)^2/4$, the ${\cal O}_{+}$ is normalizable.
Furthermore, since it is possible that the analysis in previous case is applied to any $m^2$
in the range $m_{\rm BF}^2<m^2<0$,
the chemical potential $\mu_{c}$ is investigated for more general squared mass $m^2$.
We now deal with operator of the dimension $\lambda_{+}=d/2$ before operators of
general dimensions.

In the same way in previous case $\mu_{\rm min(-)}$,
taking $m^2=d(d-2)/4$, the dimension of operator $\lambda_{+}$ (\ref{lam1}) reduces to $\lambda_{+}=d/2$.
Then the minimum eigenvalue $\mu_{\rm min(+)}$ is obtained as
\bear\label{cofmin2}
\mu_{\rm min(+)} = \frac{s_{\mu_{\rm min(+)}}(d)}{t_{\mu_{\rm min(+)}}(d)},
\eear
with
\bear
s_{\mu_{\rm min(+)}}(d) &=& \Bigg\{11d^5-33d^4-13d^3+94d^2-60d
-d\bigg (53d^{8}-266d^7-114d^6\\\nonumber
&&+2558d^5-3451d^4-4192d^3+13804d^2-12000d^1+3600\bigg)^{1/2}\Bigg\}^{1/2},\\\nonumber
t_{\mu_{\rm min(+)}}(d) &=& 2\bigg( (2d-1) (2d-3) (2d-5)\bigg)^{1/2}.
\eear
The critical value $\mu_{c}=\mu_{\rm min(+)}\thickapprox 1.890$ for $d=5$ is absolute agreement with
the numerical result in \cite{Cai:2011ky},
and $\mu_{\rm min(+)}\thickapprox 2.205$ for $d=6$,
and $\mu_{\rm min(+)}\thickapprox 2.531$ for $d=7$.

\begin{figure}[!htbp]
\begin{center}
{\includegraphics[width=12cm]{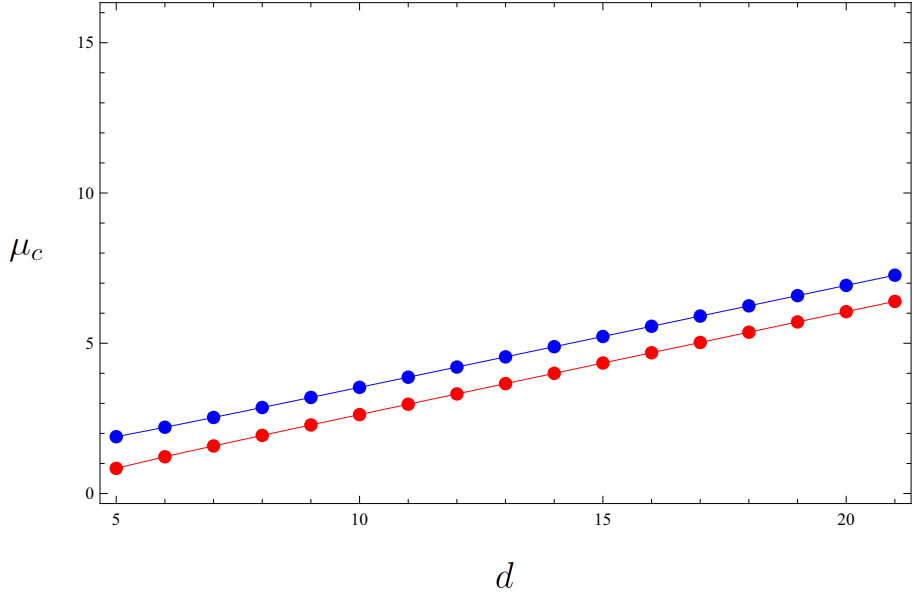}}
\end{center}
\caption{{\small The critical chemical potential $\mu_{c}$ is plotted as
the total dimension $d=5$ to $d=21$
where red is the dimension of operator ${\lambda_{-}}=(d-2)/2$ and blue $\lambda_{+}=d/2$.}}
\label{fig1}
\end{figure}

After taking the dimension of operator ${\lambda_{-}}=(d-2)/2$ and $\lambda_{+}=d/2$,
we obtain the critical chemical potential $\mu_{c}$ as the total dimension $d=5$ to $d=21$,
and so it is linearly proportional to $d$. We plot these results in Figure \ref{fig1}.

\begin{figure}[!htbp]
\begin{center}
    \begin{center}
    \includegraphics[width=7.366cm]{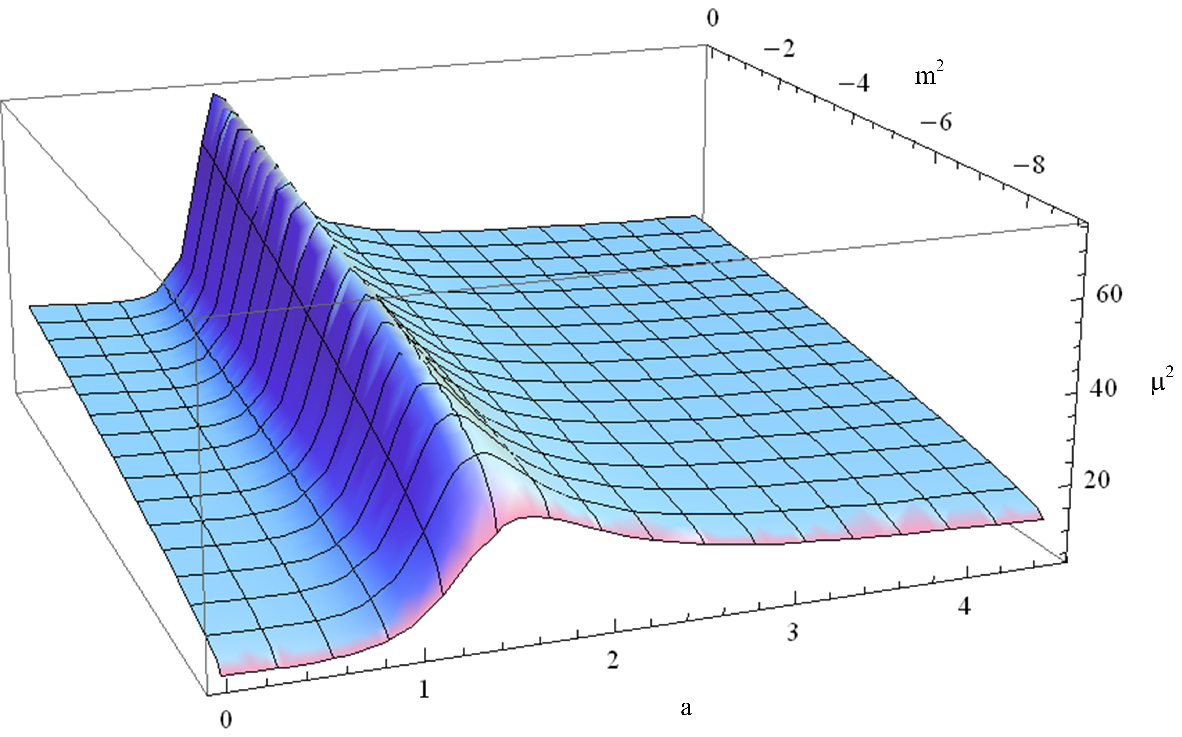}\hspace{1cm}
    \includegraphics[width=7.366cm]{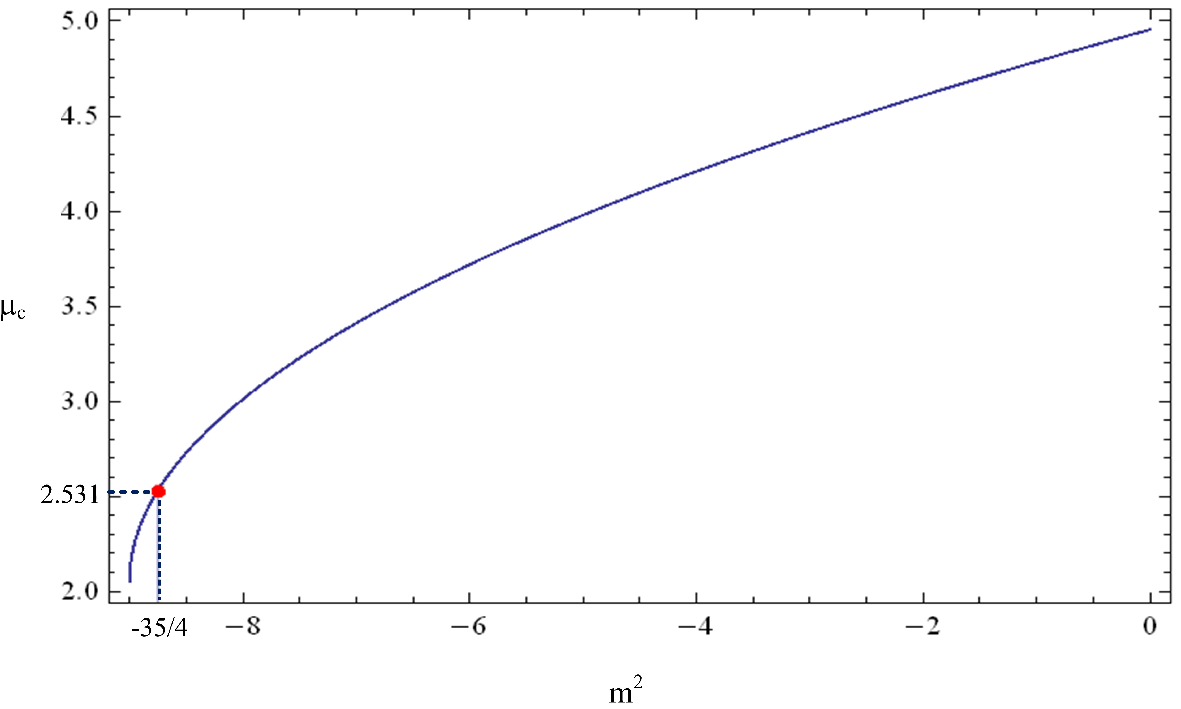}\\
    \hspace{0.8cm}(a)\hspace{7.7cm}(b)
    \end{center}
\end{center}
\caption{{\small (a) The square of chemical potential $\mu^2$ is plotted as
the constant $a$ and the square of mass $m^2$ for $d=7$.
(b) A plot of the function $\mu_{c}(m^2)$ for $d=7$.
$\mu_{c}$ has 2.531 when $m^2=-35/4$.}}
\label{fig2}
\end{figure}

Considering the operators of more general dimensions, the square of chemical potential is obtained as
\bear\label{chemipoGend}
\mu^2=\frac{s_{\mu_{m^2}}(d)}{t_{\mu_{m^2}}(d)},
\eear
with
\bear
s_{\mu_{m^2}}(d)&=&\left(\frac{2m^2+(d-1)\sqrt{(d-1)^2+4m^2}
+(d-2)(d-7)}{\sqrt{(d-1)^2+4m^2}+2(d-1)}\right.\\\nonumber
&&\hspace{0.3cm}\left.+\frac{8}{\sqrt{(d-1)^2+4m^2}+d-1}\right)a^2\\\nonumber
&&+2\left(\frac{2m^2+(d-1)\sqrt{(d-1)^2+4m^2}
+(d-1)^2}{\sqrt{(d-1)^2+4m^2}+2(d-2)}\right)a\\\nonumber
&&+\frac{2m^2+2(d-1)\sqrt{(d-1)^2+m^2}+(d-1)^2}{\sqrt{(d-1)^2+2m^2}+2(d-3)},\\\nonumber
t_{\mu_{m^2}}(d)&=& \frac{2a^2}{\sqrt{(d-1)^2+4m^2}
+d+1}+\frac{4a}{\sqrt{(d-1)^2+4m^2}+d-1}\\\nonumber
&&+\frac{1}{\sqrt{(d-1)^2+4m^2}+d-3}.
\eear
In spite of getting the explicit form of the critical potential $\mu_{c}$,
the result is not shown in this article since it is rather lengthy,
so we attempt to show the result for $d=7$ instead.
$\mu^2$ for $d=7$ yields
\bear\label{chemipo7d}
\mu^2=\frac{s_{\mu_{m^2}}(7)}{t_{\mu_{m^2}}(7)},
\eear
with
\bear
s_{\mu_{m^2}}(7)&=&\bigg\{18 m^4 + 786 m^2
+ 6396 \left(m^4+146 m^2+2124\right)\sqrt{m^2+9}\bigg\}a^2\\\nonumber
&&-2\bigg\{19 m^4+855 m^2+6804 \left(m^4+159m^2+2268\right) \sqrt{m^2+9}\bigg\}a\\\nonumber
&&+20 m^4+954 m^2+ 7776
\left(m^4+174 m^2+2592\right) \sqrt{m^2+9},\\
t_{\mu_{m^2}}(7)&=&\bigg(m^2+15+5 \sqrt{m^2+9}\bigg)a^2
-2\bigg(m^2+17+6 \sqrt{m^2+9}\bigg)a\\\nonumber
&&+m^2+21+7 \sqrt{m^2+9},
\eear
which leads to the minimum eigenvalue $\mu_{\rm min (+)}$
\bear
\mu_{\rm min (+)}= \frac{s_{\mu_{\rm min(+)}}(7)}{t_{\mu_{\rm min(+)}}(7)},
\eear
with
\bear
s_{\mu_{\rm min(+)}}(7)&=&\Bigg[m^8-22 m^6-513 m^4+(8 m^6-422 m^4)\sqrt{m^2+9}\\\nonumber
&&+m^2\bigg\{15198+6126\sqrt{m^2+9}+30\bigg(5 m^8+2907 m^6+200595 m^4\\\nonumber
&&+3778092 m^2+\sqrt{m^2+9} (178 m^6+28296m^4+882612 m^2+6781536)\\\nonumber
&&+20344608\bigg)^{1/2}-2\sqrt{m^2+9}\bigg(5 m^8+2907 m^6+200595 m^4+3778092 m^2\\\nonumber
&&+\sqrt{m^2+9} (178 m^6+28296m^4+882612 m^2+6781536)+20344608\bigg)^{1/2}\bigg\}\\\nonumber
&&-2\bigg\{37692+12564\sqrt{m^2+9}-255\bigg(5 m^8+2907 m^6+200595 m^4\\\nonumber
&&+3778092 m^2+\sqrt{m^2+9} (178 m^6+28296m^4+882612 m^2+6781536)\\\nonumber
&&+20344608\bigg)^{1/2}+83\sqrt{m^2+9}\bigg(5 m^8+2907 m^6+200595 m^4+3778092 m^2\\\nonumber
&&+\sqrt{m^2+9} (178 m^6+28296m^4+882612 m^2+6781536)+20344608\bigg)^{1/2}\bigg\}\Bigg]^{1/2},\\
t_{\mu_{\rm min(+)}}(7)&=&\sqrt{(m^2-7)(m^2-16)(m^2-27)}.
\eear
We plot the function (\ref{chemipo7d}) in Figure \ref{fig2}. (a) for $-9<m^2<0$,
which indicates that there is always the minimum value of chemical potential squared
for various $a$'s and $m^2$'s when $a\rightarrow0$.
As squared mass $m^2$ increases up to the Breitenlohner-Freedman bound
squared mass $m_{\rm BF}^2$, the critical chemical potential $\mu_{c}$ increases
(see in Figure \ref{fig2}. (b)).
When $\mu$ is very closely located near $\mu_{c}$, we have
\bear\label{fieldeq44}
\phi^{''}+
\left\{\frac{f^{'}}{f}-\left(\frac{d-4}{2n-1}-2\right)\frac{1}{z}\right\}\phi^{'}
=\frac{<{\cal O}_{\pm}>^2z^{-4n+2\lambda_{\pm}}F^2(z)}{\alpha n(2n-1)(-1)^{3n+1}2^n}
\frac{\phi}{(\phi^{'})^{2(n-1)}f},
\eear
by plugging Eq. (\ref{refield1}) into Eq. (\ref{fieldeq4}).
In such a limit, we may take $\phi(z)$ as
\bear
\phi(z)=\mu_{c}+<{\cal O}_{\pm}>\chi(z),
\eear
where the boundary condition near the tip imposes
\bear\label{bc1}
\chi(z)|_{z\rightarrow 1}=0.
\eear
Substituting in Eq. (\ref{fieldeq44}), we obtain
\bear\label{changedfieldeq44}
\chi^{''}-\frac{\bigg(2d(n-1)-2n+5\bigg)z^{d-1}+d-4}{(2n-1)(z-z^d)}\chi^{'}
=\frac{<{\cal O}_{\pm}>^{3-2n}z^{-4n+2\lambda_{\pm}}F^2(z)}{\alpha n(2n-1)(-1)^{3n+1}2^nf}
\frac{\mu_{c}}{(\chi^{'})^{2(n-1)}},
\eear
which for $n=1$ reduces to
\bear\label{fieldeq5}
\frac{d}{dz}\left[T_{1}(z)\chi^{'}\right]=-\frac{<{\cal O}_{\pm}>\mu_{c}F^2(z)}{2\alpha}
\frac{z^{2+2\lambda_{\pm}}}{z^d}
\eear
by introducing the function $T_{1}(z)$
\bear
T_{1}(z)=\frac{z^{d-1}-1}{z^{d-4}}.
\eear
Considering the operator of dimension $\lambda_{-}=(d-2)/2$ and taking $\alpha=1/4$,
the above Eq. (\ref{fieldeq5}) is obtained as
\bear
 \frac{d}{dz}\left[\frac{z^{d-1}-1}{z^{d-4}}\chi^{'}\right]
 =-2<{\cal O}_{-}>\mu_cF^2(z),
\eear
from which it follows that, integrating both sides,
\bear\label{fieldeq6}
\left.\frac{z^{d-1}-1}{z^{d-4}}\chi^{'}\right|_{0}^{1}&=&\left.\frac{\chi^{'}}{z^{d-4}}\right|_{z\rightarrow 0}
=-2<{\cal O}_{-}>\mu_c\left.z\left(\frac{a^2z^4}{5}-\frac{2az^2}{3}+1\right)\right|_{0}^{1}\\\nonumber
&=&-2<{\cal O}_{-}>\mu_c\left(\frac{a^2}{5}-\frac{2a}{3}+1\right).
\eear
$\phi(z)$ near $z=0$ is asymptotically given as
\bear\label{fieldeq66}
 \phi(z)|_{z\rightarrow 0} \thicksim \mu-\rho z^2 \thickapprox \mu_{c}+<O_{-}>
 \bigg(\chi(0)+\chi^{'}(0)z+\frac{1}{2}\chi^{''}(0)z^2+{\cal O}(z^3)\bigg),
\eear
which leads to
\bear\label{mu2}
\mu-\mu_{c}=<{\cal O}_{-}>\chi(0),
\eear
by comparing the coefficients of zeroth order in $z$ in both sides,
and from first order we can read
\bear\label{bc2}
\chi^{'}(0)=0.
\eear
After imposing two boundary conditions (\ref{bc1}) and (\ref{bc2}),
Eq. (\ref{fieldeq66}) $\chi(z)$ for $d=7$ is explicitly obtained as
\begin{align}
\chi(z)=
\begin{array}{cl}
&\frac{<{\cal O}_{-}>\mu_{c}}{90}\bigg[-\frac{36}{5}a^2(z^5-1)+120a(z-1)-\frac{3}{2}(3a^2-10a+15)\ln(z^4+1)\\
&\hspace{1.74 cm}+2(3a^2-10a+15)\ln(z^3+1)+15\sqrt{2}a\ln(z^2-\sqrt{2}z+1)\\
&\hspace{1.74 cm}-15\sqrt{2}a\ln(z^2+\sqrt{2}z+1)+4\sqrt{3}(3a^2-10a+15)\tan^{-1}(\frac{2z-1}{\sqrt{3}}z)\\
&\hspace{1.74 cm}+15(\sqrt{2}-1)\tan^{-1}(\sqrt{2}z+1)-15(\sqrt{2}+1)\tan^{-1}(\sqrt{2}z-1)\\
&\hspace{1.74 cm}+3\bigg\{3(\sqrt{2}z+1)a^2-10a\bigg\}-3\bigg\{3(\sqrt{2}z-1)a^2+10a\bigg\}\\
&\hspace{1.74 cm}-\frac{1}{2}(3a^2-10a+15)\ln(2)+30\sqrt{2}a\coth^{-1}(\sqrt{2})\\
&\hspace{1.74 cm}-\frac{1}{12}\bigg\{3(9-18\sqrt{2}+8\sqrt{3})a^2-10(9+8\sqrt{2})a+15(9-18\sqrt{2}+8\sqrt{3})\bigg\}\pi\bigg]
\end{array}.
\end{align}
Thus, from Eq. (\ref{mu2}) we get the qualitative relation between the condensation
value $<{\cal O}_{-}>$ and the chemical potential difference ($\mu-\mu_{c}$) for arbitrary dimension $d$
\bear\label{calon}
<{\cal O}_{-}>\thicksim \gamma_{-} \sqrt{\mu-\mu_{c}},
\eear
and comparing the coefficients of the $z^2$ term in (\ref{fieldeq66}), we read
the linear relation between the charge density $\rho$ and ($\mu-\mu_{c}$)
\bear
\rho\thicksim \delta_{-} (\mu-\mu_{c}),
\eear
where $\gamma_{-}$ and $\delta_{-}$ are positive constants.
For example $\gamma_{-}$ and $\delta_{-}$ for $d=5$, $d=6$, and $d=7$ are given as
\begin{align}
\gamma_{-}= \left\{
\begin{array}{cl}
&1.940~~~~~{\rm for}~d=5\\
&1.987~~~~~{\rm for}~d=6\\
&2.042~~~~~{\rm for}~d=7
\end{array}
\right.,\hspace{1.74 cm}
\delta_{-}= \left\{
\begin{array}{cl}
&2.700~~~~~{\rm for}~d=5\\
&4.050~~~~~{\rm for}~d=6\\
&5.399~~~~~{\rm for}~d=7.
\end{array}
\right.
\end{align}
Taking $\lambda_{+}=d/2$, the Eq. (\ref{fieldeq5}) is
\bear
\frac{d}{dz}\left[\frac{z^{d-1}-1}{z^{d-4}}\chi^{'}\right]=-2<{\cal O}_{+}>\mu_{c} F^2(z)z^2,
\eear
which leads to
\bear
\left.\frac{\chi^{'}}{z^{d-4}}\right|_{z\rightarrow 0}=-2<{\cal O}_{+}>\mu_{c}
\left(\frac{a^2}{7}-\frac{2a}{5}+\frac{1}{3}\right).
\eear
From following the preceding steps, we obtain
\bear\label{calop}
<{\cal O}_{+}>\thicksim \gamma_{+} \sqrt{\mu-\mu_{c}}\,,
\hspace{1cm}
\rho\thicksim \delta_{+} (\mu-\mu_{c}),
\eear
where $\gamma_{+}$ and $\delta_{+}$ are positive constants.
For example $\gamma_{+}$ and $\delta_{+}$ for $d=5$, $d=6$, and $d=7$ are given as
\begin{align}
\gamma_{+}= \left\{
\begin{array}{cl}
&1.801~~~~~{\rm for}~d=5\\
&2.099~~~~~{\rm for}~d=6\\
&2.316~~~~~{\rm for}~d=7
\end{array}
\right.,\hspace{1.74 cm}
\delta_{+}= \left\{
\begin{array}{cl}
&1.329~~~~~{\rm for}~d=5\\
&1.994~~~~~{\rm for}~d=6\\
&2.659~~~~~{\rm for}~d=7.
\end{array}
\right.
\end{align}

\begin{figure}[!htbp]
\begin{center}
    \begin{center}
    \includegraphics[width=7.366cm]{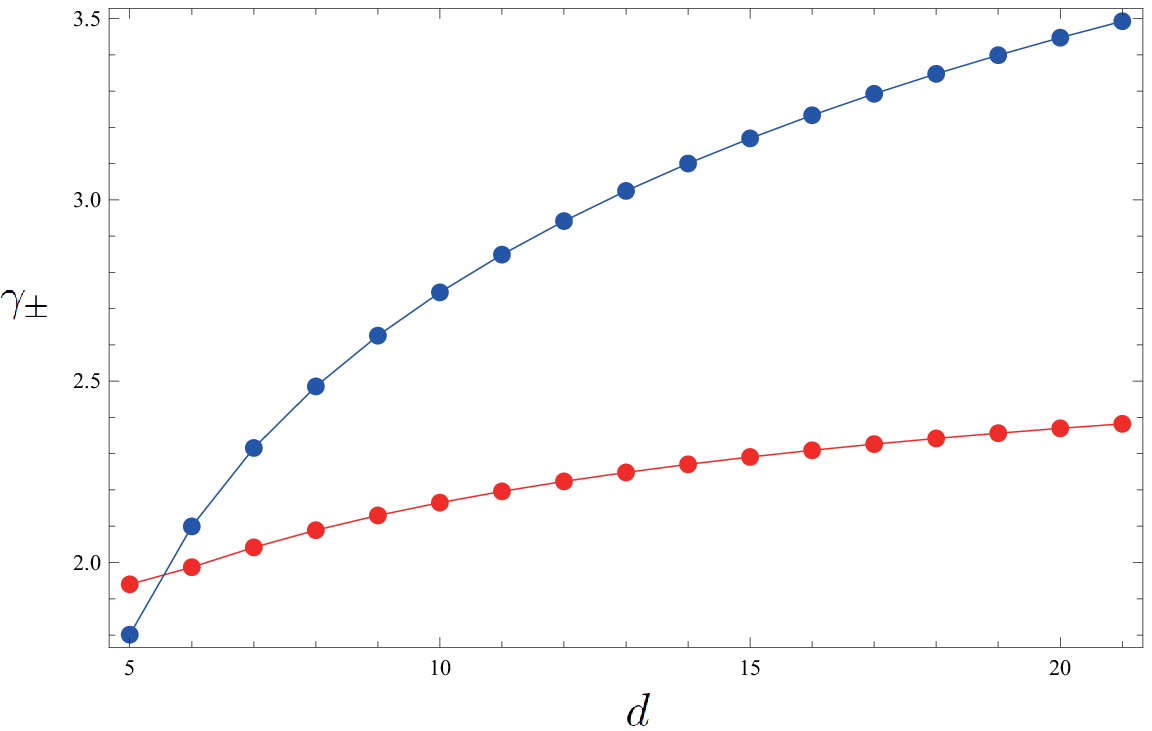}\hspace{1cm}
    \includegraphics[width=7.366cm]{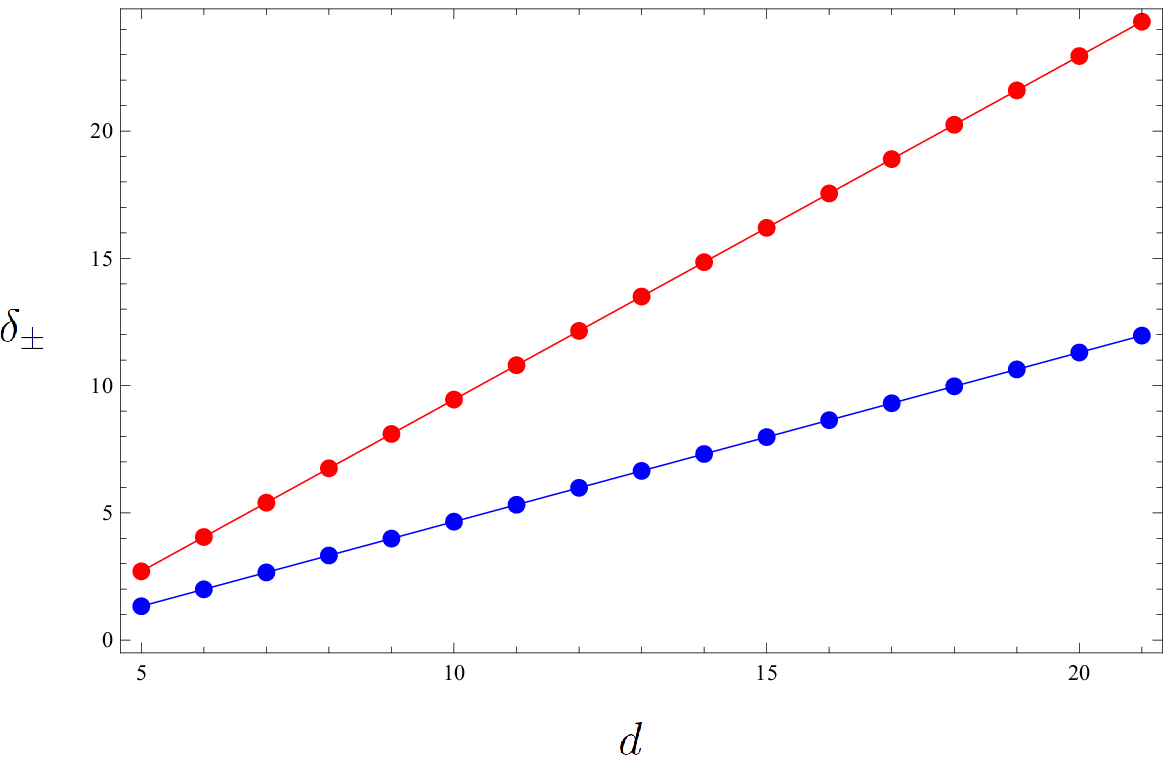}\\
    \hspace{0.8cm}(a)\hspace{7.7cm}(b)
    \end{center}
\end{center}
\caption{{\small (a) The coefficient $\gamma_{\pm}$ in Eqs. (\ref{calon}) and (\ref{calop})
is plotted as the total dimension $d=5$ to $d=21$. (b) The coefficient $\delta_{\pm}$
is plotted as $d=5$ to $d=21$. Here, red is the dimension of operator ${\lambda_{-}}=(d-2)/2$
and blue $\lambda_{+}=d/2$.}}
\label{fig3}
\end{figure}

As Figure \ref{fig3} shows, supposing the total dimension $d$ more increases than $d=5$,
the coefficient $\gamma_{+}$ in Eq. (\ref{calop}) is bigger than the coefficient
$\gamma_{-}$ Eq. (\ref{calon}), and $\delta_{\pm}$ increase linearly as $d$ grows up.

We now come back to any power of PM field $n$, and the Eq. (\ref{changedfieldeq44})
leads to
\bear\label{fieldeqn5}
\frac{d}{dz}\left[T_{n}(z)(\chi^{'})^{2n-1}\right]=
-\frac{<{\cal O}_{\pm}>^{3-2n}z^{-2n+2\lambda_{\pm}}(z^d-1)^{2n-2}F^2(z)}
{\alpha n (-1)^{3n+1}2^n}\mu_{c},
\eear
with
\bear
T_{n}(z)=\frac{(z^{d-1}-1)^{2n-1}}{z^{d-4}}.
\eear
Then the condensation value $<{\cal O}_{\pm}>$ and the charge density $\rho$ are qualitatively
\bear
<{\cal O}_{\pm}>\thicksim \xi_{\pm} (\mu-\mu_{c})^{\frac{1}{4-2n}}\,,
\hspace{1cm}
\rho\thicksim \zeta_{\pm} (\mu-\mu_{c})^{\frac{1}{2-n}},
\eear
where $\xi_{\pm}$ and $\zeta_{\pm}$ are positive constants. It implies that
the critical exponent of condensation operator can be changed into $1/(4-2n)$ for various $n$ unlike that
of Maxwell field.

\section{Conclusion}
Previous work on the analytical behavior of the holographic
superconductors in five-dimensional AdS soliton spacetime
\cite{Cai:2011ky} have found that the critical exponent of
condensation operator is 1/2, and the charge density is linearly
depending on the chemical potential. We also get the same results in
higher-dimensional cases. However, the critical exponent of
condensation operator is changed into $1/(4-2n)$ in the context of
the Einstein-Power-Maxwell field theory, and the charge density is
proportional to the chemical potential to the power of $1/(2-n)$. In
addition, since analytical calculations in Eqs. (\ref{calon}) and
(\ref{calop}) indicate AdS soliton background is unstable below the
threshold value $\mu_{c}$ but are stable above this value, they
may play the role of higher-dimensional insulator and superconductor
in the dual field theory, respectively, as in the case of
$5$-dimensional AdS soliton \cite{Nishioka:2009zj, Cai:2011ky,
Peng:2011gh}.

\end{document}